# FIRST LIGHT FROM THE DOME C (ANTARCTICA) OF A PHASE KNIFE STELLAR CORONAGRAPH

Geraldine Guerri[1], Lyu Abe[1], Jean-Baptiste Daban[1], Eric Aristidi[1],
Philippe Bendjoya[1], Jean-Pierre Rivet[2] and Farrokh Vakili[1]

*February 12, 2009*

ABSTRACT

We report on the first daytime on-sky results of a Phase Knife stellar Coronagraph operated in the visible from the French-Italian Concordia station at Dome C of Antarctica. This site has proven in the last few years to offer excellent atmospheric seeing conditions for high spatial resolution observations.

The coronagraphic performances obtained from laboratory experiments and numerical models have been compared with those measured from daytime on-sky data recorded on bright single and multiple stars: Canopus (HD 45348), and α Centauri (HD 128620J). No correction system was used (adaptive optics or tip-tilt mirror) so that atmospheric turbulence alone defines the image quality, and thus the coronagraphic performances. Moreover, the experiment could not run under optimal operational conditions due to hardware/software problems.

Satisfactory results have been obtained: broad band total rejection exceeding 15 were attained in the visible. This first day-time observation campaign yields an experimental feedback on how to improve the instrument to get optimal performances during future night-time observation runs.

## 1    INTRODUCTION

Since the discovery by Mayor & Queloz (1995) of the first extrasolar planet orbiting a solar-like star, the competition to obtain direct images of an exoplanet has been launched. However, the huge brightness ratio between the star and its orbiting planet ($10^9$ to $10^6$ depending on the spectral domain and on the planet's nature) requires the use of new high dynamic range imaging instruments. Stellar coronagraphy is a promising technique to overcome this brightness ratio, as recently confirmed by Neuhauser *et al.* (2007) with the direct detection of an exoplanet in the system γ Cephei, and by Boccaletti *et al.* (2008) with the accurate photometric measurement of the close companion AB Doradus C. In this context, several coronagraphic concepts have been proposed in the literature: Roddier & Roddier phase mask coronagraph (Roddier & Roddier 1997), achromatic interferometric coronagraph (Gay *et al.* 1997), four-quadrant phase mask coronagraph (Rouan *et al.* 2000, Mawet *et al.* 2006), achromatic phase knife coronagraph (Abe *et al.* 2001, Abe 2002), or pupil apodized Lyot coronagraph (Aime *et al.* 2002, Soummer *et al.* 2003).

---

[1] Laboratoire A.H. Fizeau, Université de Nice Sophia-Antipolis, CNRS, Parc Valrose, 06108 Nice Cedex 02, France

[2] Laboratoire Cassiopée, Université de Nice Sophia-Antipolis, CNRS, Observatoire de la Côte d'Azur, B.P. 4229, 06304 Nice Cedex 04, France. *Corresponding author* : Jean-Pierre.Rivet@oca.eu

The main limitations of ground-based coronagraphs are the atmospheric turbulence and the performances of adaptive optics systems: currently, good on-sky coronagraphic performances are difficult to obtain, even with the best adaptive optics system (Boccaletti *et al.* 2004). Choosing an observing site with the best possible atmospheric quality is thus of crucial importance. The Dome C site in Antarctica provides excellent environmental conditions for astronomical observations thanks to its atmosphere stability and, consequently, its low rate of turbulence. That is why a ground-based telescope equipped with an adaptive optics system and installed at Dome C should be very competitive for stellar coronagraphy and high dynamics imaging.

We describe in this article the CORONA experiment (for CORONagraph in Antarctica), which is a first preliminary step towards the achievement of a future high efficiency coronagraphic facility in Antarctica. It involves a precursory instrument from a technical point of view (regarding the coronagraphic device, the acquisition system and the operation in extreme conditions). The goal of this work is first to assess the feasibility of stellar coronagraphy at Dome C with a telescope having central obstruction, and second, to yield experimental feedback for a future more ambitious instrument, equipped with a tip-tilt corrector and an adaptive optics system.. Indeed this experiment is based on a "Lucky Imaging" approach (Baldwin *et al.* 2001) where a simple tip-tilt would probably significantly increase the instrument performance. We first recall some basic facts on the French-Italian Antarctic station Concordia, where the CORONA instrument has been set up (Section 2). Then, we describe the instrument itself (Section 3) and the preliminary laboratory tests (Section 4) that were performed in France before the first summer observing campaign, the results of which are presented in Section 5 and discussed in Section 6. Finally, the experimental feedback is discussed and some conclusions are drawn (Section 7).

## 2 THE CONCORDIA STATION

The Concordia station was constructed by the French and Italian polar institutes (*Institut Paul-Emile Victor* – IPEV, and *Programma Nazionale di Ricerche in Antartide* - PNRA). It is located on the Dome C site on the Antarctic plateau (75°06′ S, 123°20′ E), at an elevation of 3233 m, which corresponds to an air pressure met around 3800 m at more equatorial latitudes. The facility is available for summer activities since December 1997 whilst winter activities began in 2005 only. Three entire winter-overs have been achieved since the first one, in 2005. The first results of the site testing depict Concordia as an exceptional site for observational astrophysics. The daytime seeing measurements (Aristidi *et al.* 2003 and 2005b) were based on DIfferential Motion Monitor (DIMM) data, during two summer campaigns (3 months each) in April 2003 and May 2004. They found a median seeing of 0.54″ (arc seconds) and a median isoplanatic angle of 6.8″ at more than 8.5 m above the ground level. The following nighttime measurement campaign (Agabi *et al.* 2006, Trinquet *et al.* 2008), however, showed that the median seeing measured 30 m above the ground level was equal to 0.36″, with exceptionally low refractive index structure constant and wind speed profiles, challenging top-ranked observing sites like the Mauna-Kea volcano in Hawaii or the Cerro Paranal in Chile (see *e.g.* Sarazin & Tokovinin, 2001). These results strongly encourage the astronomical community to consider Dome C as a possible site for the construction of a new large observatory mainly dedicated to high dynamic range imaging observations and interferometric arrays (Vakili *et al.* 2005). In this context, and pending the installation of instruments at larger scale, we present patrol experiments of a visible stellar coronagraph which have been conducted at Dome C, in parallel to the site qualification.

## 3 THE CORONA INSTRUMENT

CORONA is a compact four-quadrant phase mask stellar coronagraph with reduced chromatism, coupled to a 14-inch (355 mm) telescope. The coronagraph has been designed to work in the visible and to withstand Antarctic conditions. The telescope itself has been modified accordingly.

### 3.1 Four-quadrant Phase Mask Coronagraphs

Four-quadrant phase mask coronagraphs (Rouan *et al.*, 2000) are potentially good candidates for stellar coronagraphy in severe environmental conditions. In Figure 1, we sketch their basic principle: the central optical component is a phase mask divided into four equal quadrants, each one being supposed to shift the phase of an incoming light by respectively 0, π, 0, and π. Ideally, the phase mask should be spatially infinite, and the phase shifts should be achromatic, without transmission loss, with sharp transitions at the limits of the quadrants. The phase mask has to be placed in the focal plane of the telescope, and the central point of the mask should be exactly placed on the telescope's optical axis. After the phase mask, a relay lens produces a real image of the telescope pupil. It has been proven theoretically and verified experimentally (Abe *et al.* 2003), that the effect of the phase mask is to reject all of the light from an on-axis unresolved star outside of the geometrical image of the pupil, provided that the pupil is perfectly circular, without any central obscuration, that the phase mask is ideal, and that no atmospheric turbulence occurs. To eliminate this unwanted light, an iris diaphragm called "Lyot stop", slightly smaller than the geometrical image of the pupil, is placed in the relayed pupil plane. Of course, the light from the off-axis source to be studied is preserved, or at least less affected. Finally, a last lens focuses the remaining light onto the camera. In this final focal plane, the residual light from the on-axis star would completely vanish in ideal conditions. However, with a non-ideal mask, the final coronagraphic image of an on-axis source is not totally dark, even without atmospheric turbulence: the residual halo has a particular light distribution similar to "butterfly's wings" (see Figure 1, bottom-right image). The intensity level and exact shape of the halo mostly depends on the actual finite size of the coronagraphic mask (see, e.g. Abe *et al.* 2003), and also on other defects of the mask (phase errors, chromaticity, homogeneity). In addition, if the incoming stellar light is distorted by the atmospheric turbulence and/or by static optical aberrations in the telescope, then, the residual halo has a more complex structure at a higher photometric level. The performances of stellar coronagraphs are thus very sensitive to atmospheric turbulence, and it is necessary to implement the experiment on a site with very good seeing conditions, especially if no adaptive optics is available.

### 3.2 The phase mask design

The design of the four-quadrant phase mask is one of the most critical issues in the realization process of the coronagraph. It must be optically as close as possible to the ideal case, and be as compact, robust and reliable as possible to be compatible with the severe constraints imposed by a full on-sky observation cycle in Antarctica. The technical solution chosen for the phase mask has already been successfully tested on a first prototype which was designed for European weather conditions and settled at the focus of a 50 cm refractor at the Observatoire de Nice (Abe *et al.* 2007). Further research and development has been completed to adapt the phase mask to the observation conditions encountered at Dome C.

The mask is an assembly of two crossed pairs of phase plates, called "phase knives" (by analogy to the Foucault's optical knife test). Both phase plates of each phase knife have the same thickness (99 µm) but different refraction indices and dispersion. As shown in Figure 2, the two phase knives are sandwiched between two significantly thicker glass plates. The optical nature of the glasses and the thickness of the plates are carefully chosen to yield phase shifts close to the desired values, with reduced chromatic variations over the spectral domain of interest, *i.e.* the visible.

This kind of complex optical components requires multiple assembly steps involving both molecular bonding and gluing. To avoid penetration of humidity, an external cladding is necessary: the coronagraphic mask is covered with a flexible coating which withstands the extreme -70°C temperatures at Dome C. Finally, the component had to be dried for a few weeks before being ready for insertion in the coronagraph.

### 3.3 The coronagraphic setup

Figure 3 sketches the optical scheme of the CORONA instrument. The coronagraph itself is mounted on the side of the telescope. The output beam of the telescope passes through a beam-splitter which transmits 80% of the light to a monitoring camera, a standard "Watec" camera (not represented in Figure 3). Unfortunately, for the first observation run reported on in this article, the monitoring camera could not be used due to software problems. Thus, neither auto-guiding nor run-time image selection was available (see Section 5). The remaining 20% of the light are reflected through the coronagraph to the science camera, a "PCO Pixelfly" visible camera (360 nm to 640nm). The coronagraphic phase mask is placed at the focus of the telescope after the beam splitter. After the phase mask, the relay lens $L_1$ ($f_1 = 150$ mm) re-images the entrance pupil. The Lyot stop is placed at the relayed pupil plane. Then, two imaging lenses $L_2$ ($f_2 = 75$ mm) and $L_3$ ($f_3 = 50$ mm) can be switched to image respectively the focal image or the pupil image onto the science camera. Indeed, the pupil image is required for the preliminary optical alignment procedure.

The science camera is placed into an insulated and thermally controlled box. The typical temperature inside the box is thus around -15° C. However, in case of power failure, the temperature inside the box can drop down to -65°C or even lower. Thus, the coronagraphic mask (which is designed to operate at temperatures around -15° C) must be able to withstand those extremely low temperatures without getting damaged, and to recover its optical properties when heated back to its operating temperature.

### 3.4 The telescope

CORONA's telescope is a 14-inch (355 mm) Schmidt-Cassegrain telescope with a Barlow lens (equivalent focal length 24406 mm). The telescope is placed on an equatorial "AstroPhysics 1200" mount. The mount is bolted on a massive wooden pillar (Aristidi *et al.* 2005b).

The mechanical structure of the telescope has been modified to withstand Antarctic conditions. The original aluminium alloy optical tube has been replaced by an Invar$^{TM}$ tube (the thermal expansion coefficient for Invar$^{TM}$ is more than ten times lower than for aluminium alloys). The two mirrors of the telescope are also glued to their holder according to a special procedure. Finally, all moving mechanical parts are dry-cleaned then lubricated with a low temperature grease which remains viscous down to -80°C.

It is known that the telescope's central obscuration would significantly alter the coronagraphic nulling efficiency (Riaud *et al.* 2001). However, Lloyd *et al.* (2003) proposed several entrance pupil geometries allowing better coronagraphic performance when a central obscuration is present, but of course at the expense of a loss of transmission. The new entrance pupil is composed of four equal holes surrounding the central obscuration (see Figure 4). The pupil transmission is thus reduced to 42.7% of the full aperture transmission.

The shape of the corresponding Lyot stop is obtained from the geometry of the entrance pupil through: *(i)* a global scaling according to magnification factor of the optics, *(ii)* a reduction of 30% of the holes' diameter to improve rejection. Note that Serabyn *et al.* (2006)

proposed to use a single off-axis circular sub-aperture, which is an alternative solution to the same problem.

## 4 THE LABORATORY TESTS

To assess its ability to withstand low temperatures in case of temporary failure of the thermal control facility, the coronagraphic phase mask was first brought to Dome C in November 2004, and exposed to the local atmospheric conditions. This did not reveal any visible mechanical alteration. To verify that its optical properties had not been altered either, the phase mask was sent back to France, and laboratory tested as described below.

### *4.1 Test with a single circular aperture*

The purpose of this first set of tests was to commission the phase mask component itself. A Helium-Neon laser source (wavelength = 633 nm) has been used together with a set of standard optical components so as to produce a clean converging beam similar to the one produced by the CORONA telescope (same numerical aperture). The shape of pupil used in this first set of validation experiments was a single circular aperture. The phase mask was settled in the focal plane and pupil images were recorded for a first visual qualitative check of the phase mask efficiency.

Figure 5 displays both recorded (frames **c** and **d**) and simulated (frames **a** and **b**) pupil images corresponding to the following situations: the source is off-axis, but its focal image lies on the edge of one single phase knife (frames **a** and **c**); the source is on-axis, that is, its focal image lies at the intersection of the edges of both phase knives (frames **b** and **d**). Qualitatively, the recorded and simulated images are in agreement.

For a more quantitative comparison, we measured the coronagraphic nulling performances. To achieve this goal, we have added a Lyot stop in the relayed pupil plane, the circular hole of which is 30% smaller than the geometrical entrance pupil image.

Figure 6 shows the recorded reference (non coronagraphic) image of an off-axis point source (frame **a**), the coronagraphic residual image of the same point source, when on-axis (frame **b**; intensity scale multiplied by 1000), and their average radial profiles with a logarithmic intensity scale (frame **c**).

To evaluate the coronagraphic nulling performances, we measure the following quantities (see *e.g.* Abe *et al.*, 2003):

- *Extinction ratio*: Ratio of the peak intensity in the reference non-coronagraphic images (source off-axis), to that of coronagraphic images (source on-axis).
- *Energy rejection ratio*: Ratio of the total transmitted energy in the reference non-coronagraphic images, to that of coronagraphic images. This quantity can be measured in the pupil plane, or in the image plane.

Note that these two quantities are greater than one. The larger they are, the better the coronagraph is. No simple relation exists between these quantities, and both need to be measured independently.

In this first set of laboratory experiments (single circular entrance pupil, monochromatic light), the extinction ratio has been found to fluctuate around 900, with peak values reaching 1100. The rejection ratio (measured in the coronagraphic focal plane) is around 500 with peaks at 700.

### *4.2 Test with four circular sub-apertures*

To be closer to the actual geometry of the CORONA instrument, we have conducted a second set of laboratory experiments, similar to the one described in Section 4.1, but with a four-hole entrance pupil. The Lyot stop for this set of experiments has the same shape as the image of the entrance pupil, but with the four holes 15% smaller holes.

Figure 7 displays the pupil images recorded with the same experimental test bench, in the following three situations: the source is fully off-axis (frame **a**); the source is off-axis, but its focal image lies on the edge of one single phase knife (frame **b**); the source is on-axis (frames **c**). As expected, the four-hole version of the coronagraph works the same way as its single-hole counterpart: the effect of the phase mask is to reject the light from an on-axis source outside of the geometrical images of the four holes in the entrance mask.

Figure 8 is strictly analogous to Figure 6, but corresponds to the four-hole case. The reference focal image of an off-axis source (frame **a**) as well as the coronagraphic residue of an on-axis source (frame **b**; intensity scale multiplied by 1000) display more complex structures, due to superimposed interference patterns. The radial profiles in frame **c** are less clean than in the single aperture case (Figure 6). This is by no means surprising since the focal images (even the non-coronagraphic one) are not expected to be axis-symmetric with a four-hole entrance pupil.

In this second configuration (four-hole pupil, monochromatic light), the extinction ratio was measured around 600, with peaks at 650. The rejection ratio remains below 100.

## 5   ON-SKY TESTS

CORONA's first light at Dome C was obtained on December 5th, 2005 at 11h AM local time (UTC+ 8 hours). Figure 9 shows the CORONA instrument on its wooden pillar, ready for observations.

### 5.1   *Preliminary operations*

A good optical alignment of the telescope and of the coronagraph is essential to obtain the best performances. To correct for the optical misalignments that were likely to occur during the transportation from France to Antarctica, an optical bench was set up in the Concordia laboratory at the beginning of the summer campaign in November 2005. A collimated monochromatic beam was produced with a pinhole lit by a laser at the focus of a 16-inch (406 mm) Schmidt-Cassegrain telescope. This collimated beam was fed to the CORONA instrument and used to tune accurately the positions of the phase mask and of the Lyot stop.

CORONA was installed near the two wooden platforms of the Concordia Observatory at an elevation of 1.5 m above the ground. These platforms are located 300 m away from the Concordia station, in South-West direction to avoid the turbulence generated by the local power plant (Aristidi *et al.* 2005a). CORONA was operated from an igloo hut located 10 m away from the telescope.

Polar alignment was made using the Bigourdan method on solar spots, and fine tuning was made on a bright star (HD 45348) during the observations. The collimation of the telescope could not be performed perfectly, because of the high level of daytime sky background. Indeed, even with the brightest visible star (Canopus), the telescope could not be sufficiently defocused without loosing the star image.

### 5.2   *The stars*

For this first daytime on-sky observation campaign, bright single and multiple stars have been chosen:
  a) HD 45348 ("α Carinae" or "Canopus"), a bright single star (spectral type F0, magnitude -0.72). This star has been used for polar alignments and optical fine tuning, since it is ten times brighter than the daytime sky background.

b) HD 128620J ("α Centauri"), a triple star whose faintest component ("α Centauri C", also known as "Proxima Centauri") has magnitude 11 and is too faint to be detected by daytime with our instrument. The A component has the G2 spectral type. The magnitudes of the A and B components are -0.01 and 1.33, with an angular separation of 14.1″.

*5.3      The data reduction method*

Three different kinds of data sets have been taken for each observing run: non coronagraphic (reference) sets, with star off-axis, coronagraphic sets and sky background sets. Each data set is a sequence of 374 snapshots. The exposure time for each snapshot (10 milliseconds) have been chosen in order to freeze the effects of both atmospheric turbulence and mechanical vibrations. The size of each snapshot is 640×480 pixels, which covers 2.6′×2′ on the sky.

For several reasons, the standard astronomical data reduction algorithms could not be used directly. First of all, the geometry of the entrance pupil mask with four equal circular holes (see Figure 4) leads to complicated diffraction patterns, even for non-coronagraphic images (star off-axis). Indeed, the "point spread function" (PSF), *i.e.* the focal image of a single star, is far from being axis-symmetric, even in perfect atmospheric conditions. In addition, when the star is on-axis, the four quadrant structure of the coronagraph introduces further complications in the structure of the residual image (see for example Figure 8 for turbulence-free laboratory images of a point source off- and on-axis). Real on-sky images are of course even more complicated. Consequently, the standard algorithms to fit the photometric profile of star images with a Gaussian profile fail to converge, or converge toward erroneous values. Thus, the seeing parameter (usually defined as the Full Width at Half Maximum of the best fitting Gaussian profile) could not be reliably measured, and the positions of the stellar images were difficult to measure accurately.

Furthermore, neither tip-tilt corrector nor automatic guiding was available for these preliminary observing runs. So, the star image on the camera sensor was randomly fluctuating around its mean position, due to the atmospheric turbulence, or to wind-induced mechanical vibrations (several pixels *rms*). In addition, some residual flexions in the mechanical structure of the coronagraph could not be ruled out. As a consequence, the cross-shaped shadow of the phase mask was also slightly moving on the CCD target (about one pixel *rms*). Thus, aligning the snapshots in a data set to obtain a long exposure (the average or median of all the snapshots in a dataset after proper alignment) required to choose which structure was to be aligned (the shadow of the mask, or the stellar image). For sky background or coronagraphic datasets, we chose to align the shadow of the mask. A special alignment algorithm had thus to be designed. Reliable median and average sky background images and coronagraphic long exposures could thus be obtained.

Note that the images were recorded by daytime. Thus, the sky background level is high and some faint but relevant structures (stellar companions or coronagraphic residuals) hardly emerged out of the background noise. The background level could not be fully compensated for by median sky image subtraction, since the sky brightness is likely to vary rapidly with time (coronagraphic and non-coronagraphic images could not be recorded simultaneously). The residual background level, after median sky image subtraction, was eliminated by a method inspired from the well-known aperture photometry algorithm, but modified to incorporate the fact that the background level is different for pixels inside and outside the cross-shaped shadow of the phase mask. This method provided for a correct estimate of the photometric quantities (total flux and peak intensity) needed to compute rejection and extinction ratios.

For some data sets, the best coronagraphic snapshots had stellar residues falling at the level of the background noise. For those data sets, our ability to estimate rejection or extinction ratios is limited by the background noise. To be more quantitative, we have estimated, for each coronagraphic dataset, the overall noise affecting the photometric measurements by the following procedure: for each of the 374 snapshots in a data set, we have applied our modified aperture photometry method to a position in the field where no stellar light is expected (far from the star, companion, or stellar residuals). The resulting 374 values of course fluctuated around zero, with a standard deviation $\sigma$ we could easily compute. This standard deviation $\sigma$ served as an estimate of the noise level on the corresponding photometric variable. A coronagraphic stellar residual was considered to emerge out of the noise if its photometric value (total flux or peak intensity) was three times larger than the corresponding standard deviation $\sigma$. Otherwise, the value was not taken into account in the statistics. Our rejection and extinction ratio evaluations are thus conservative.

*5.4  The results*

As in laboratory experiments (see Section 4), we have measured the two standard quantities used to assess the nulling performances of a coronagraph: the *rejection ratio* (ratio of the total transmitted energy with the star off-axis, to the same quantity with the same star on-axis), and the *extinction ratio* (ratio of the peak intensity in the image with the star off-axis, to the same quantity with the same star on-axis). For each coronagraphic data set, these measures were performed on all individual snapshots, and also on a long exposure obtained by adding up the 10 best snapshots.

Table 1 sums up the measured coronagraphic performances for the first daytime observation campaign at Dome C. The rejection ratio reached 15 on an individual snapshot and only 11 on a 10-image long exposure (for HD 45348, on December 11$^{th}$, 2005). The best extinction ratio we could obtain is 17 on an individual snapshot, and 24 on a 10-image long exposure.

Figure 10 displays the best coronagraphic snapshots for HD 45348 (frame **a**) and HD 128620J (frame **c**), together with the reference non-coronagraphic images (frames **b** and **d** respectively) for visual comparison. Note that the photometric scale is different for the coronagraphic images (frames **a** and **c**) than for the reference images (frames **b** and **d**). For the double star HD 128620J, the companion emerges clearly on the coronagraphic frame **c**.

Coronagraphic long exposure images are obtained by averaging the best selected snapshots. As an example, Figure 11-(a) shows a 3D intensity plot for the coronagraphic long exposure of HD 45348 obtained with data recorded on December 11$^{th}$ 2005: the 700 best coronagraphic snapshots have been averaged. For visual comparison, a long exposure non-coronagraphic image of this object is shown in Figure 11-(b). It has been obtained by averaging a complete data set of reference images.

Figure 12-(a) displays the coronagraphic long exposure image for HD 128620J: it is the average of the 800 best coronagraphic snapshots recorded on December 11$^{th}$ 2005. The reference non-coronagraphic long exposure image shown in Figure 12-(b) is the average of a whole reference data set.

6   DISCUSSION

The rejection ratios (15 on the best snapshot, 11 on a 10-images long exposure) and extinction ratios (17 on the best snapshot, 24 on a 10-images long exposure) we have obtained during the first observation campaign of the CORONA instrument may seem modest when compared to laboratory results described in Section 4. However, the latter have been obtained with a monochromatic light, and without turbulence, whereas the former have been obtained over a relatively large spectral domain only limited by the sensitivity of the camera (360 nm to 640 nm), and in presence of atmospheric turbulence. The turbulence, although small at Dome C, appears to be the limiting factor for the coronagraphic efficiency. Thus, our on-sky results should be not be compared to our laboratory results, but to other on-sky results. Our results are of the same order of magnitude as those obtained by Boccaletti *et al.* (2004) on the ESO Very Large Telescope (VLT) with the NACO near-infrared camera and adaptative optics.

Several facts have to be taken into account to correctly interpret the meaning of the observation results reported in 5.4. The images were recorded during daytime, with a bright sky background. Some good coronagraphic images had to be removed from the statistics, because their coronagraphic residuals did note emerge sufficiently above the residual noise. This problem should not occur with night observations.

The coronagraphic phase mask is not perfectly achromatic. The phase shift departs from its ideal value $\pi$, especially below 400 nm. Taking into account the actual sensitivity of the science camera and the estimated transmission of the optical train, the broadband rejection ratio is not theoretically expected better than 67 for a solar-type source, even without turbulence. However, this theoretically expected rejection ratio should rise above 2000 if an UV-rejecting filter removing radiations with wavelengths smaller than 420 nm was added.

Stellar coronagraphs with good close sensing capabilities are extremely sensitive to the tip-tilt component of the atmospheric turbulence and to tracking defects. Neither adaptive optics nor tip-tilt corrector, nor auto-tracking facility was available on CORONA. This hampered severely the expectable performances.

Even with good seeing conditions, the CORONA instrument is not diffraction limited. Aberrations appear in the telescope and their origin needs to be investigated.

7   CONCLUSION

We have presented the first light results from the Dome C Concordia station (Antarctica) for an achromatic phase knife coronagraph. It should be underlined that this also represents the first on-sky results for a stellar coronagraph in Antarctica, and the first on-sky validation for the four-hole entrance mask strategy proposed by Lloyd *et al.* (2003) to address the issue of fitting a stellar coronagraph onto a reflective telescope with central obstruction. The rejection ratio of 15 measured by daytime over a relatively broad spectral band (360 nm to 640 nm, full width at half-maximum) is very encouraging even though they sound modest. In addition, these first on-sky tests led to some important technical feedbacks that will benefit to the future Dome C instruments.

During the winter, temperatures can drop down to values as low as -80°C. From the mechanical point of view, the telescope has been modified to withstand very low temperatures. However, some strong aberrations appear at winter temperatures. Indeed, when CORONA was tested during the 2006 winterover (temperature close to -65°C.), the images of the target star Sirius revealed strong aberration, with a 15 arc seconds triangle-shaped image. This aberration is probably caused by thermally induced mechanical tensions on the telescope mirrors. The coronagraph could not be operated in these conditions, and thus, no nighttime images are presently available. CORONA has been sent back to France to be modified according to the experimental feedback of the first daytime campaign:
   a) The mechanical design of the primary mirror mounting will be improved to withstand polar winter temperatures without excessive flexions and misalignments.
   b) An automatic star tracking system should be added, using the monitoring camera data to compensate for slow drifts.
   c) The coronagraphic performances would benefit from the installation of a residual tip-tilt corrector.

d) An image-selection algorithm will be implemented, using the monitoring camera images, as described in Abe *et al.* (2007).
e) An UV-rejecting filter should be inserted, to improve the coronagraph efficiency (see Section 6).

These validation observations are mostly important for the future development of ambitious astronomical long term programs at Dome C. These programs will include both high spatial resolution and high contrast instruments such as coronagraphs and nulling interferometers.

On one hand, daytime site-seeing monitoring (Aristidi *et al.* 2005b) has proved Dome C to offer exceptional atmospheric conditions during the polar summer. On the other hand, the extremely low atmospheric water content at Dome C (the average precipitable water vapour measured in summer at Dome C by Valenziano (2005) is 0.6 mm) is largely better than any known observatory such as Mauna Kea or Paranal. Besides, observations from Antarctica can benefit from enhanced duty-cycle observations and even continuous exposures of several 24h periods to image circumpolar objects. These perspectives therefore justify to progressively improving the CORONA pilot experiment, and possibly, to install it on the IRAIT 80 cm telescope (Tosti *et al.* 2006) planned to be soon operated at Dome C in 2008-2009.

**Acknowledgements:**
The authors would like to thank G.Greiss (Sud-Est Optique de Précision, France) for manufacturing the achromatic phase mask, F.Valbousquet (Optique et Vision, France) for preparing the telescope, A. Robini who manufactured the mechanical parts of the coronagraphic bench, and F. Jeanneaux, C. Combier and K. Agabi for their help.
The authors are also grateful to the staff of the summer campaign at Concordia, and to the polar institutes IPEV and PNRA for the whole logistic support and the transport.
The experiments reported in this article have been supported by the district of Provence Alpes Côte d'Azur (PACA).
G. Guerri is grateful to CNRS and the PACA district for supporting her PhD thesis.

# Figures

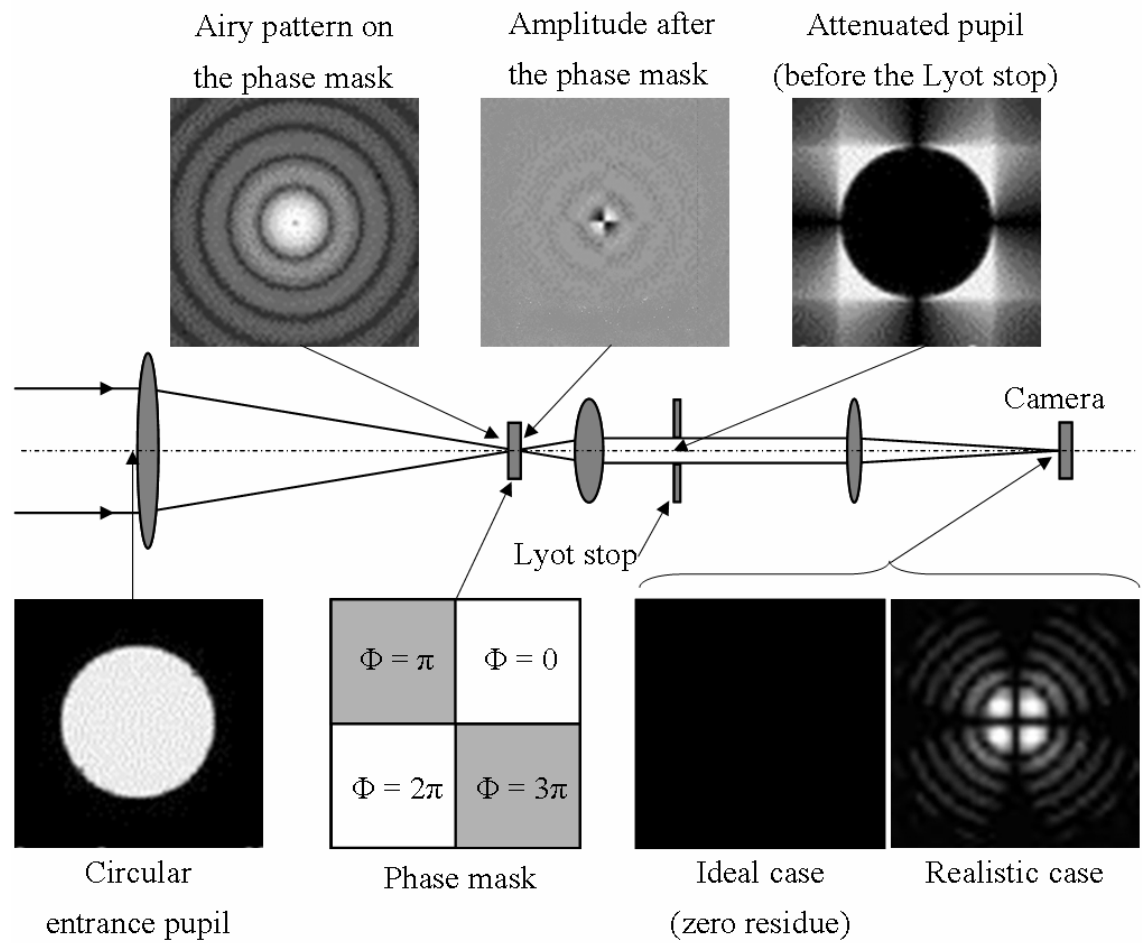

**Figure 1:** Principle of the phase knife coronagraph. The contrast has been boosted on the lower right image (realistic case) to make the butterfly-like structure visible.

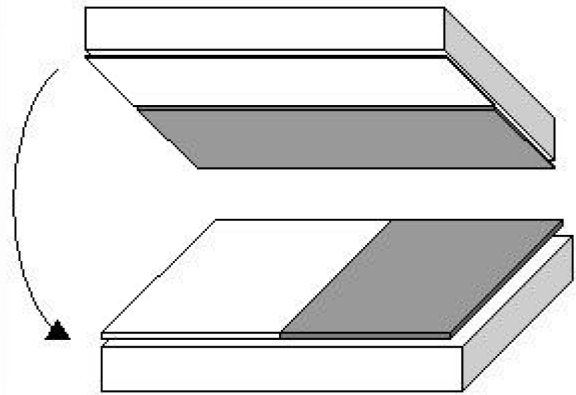

**Figure 2:** The optical assembly of the APKC coronagraphic mask: two 99 μm phase knives are stacked between two 6 mm thick glass plates.

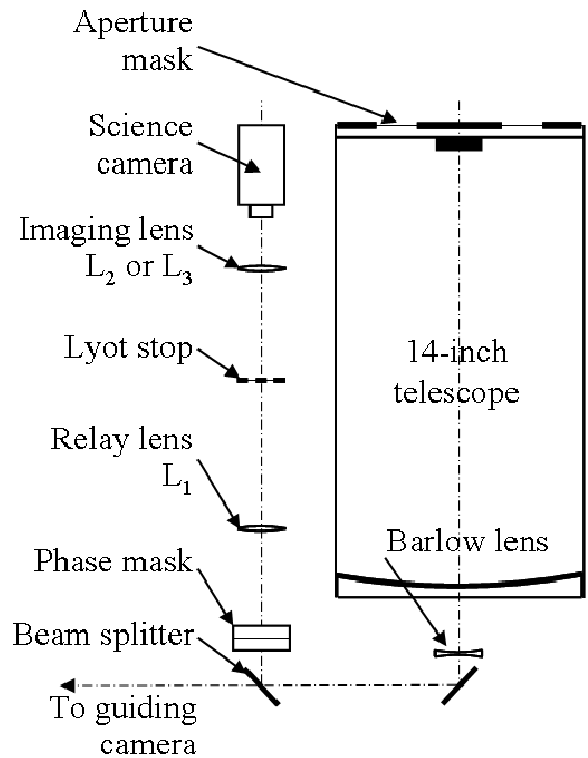

**Figure 3:** CORONA's optical setup. The 14" Schmidt-Cassegrain telescope with its four-hole aperture mask and its Barlow lens delivers a converging light beam to the coronagraph. This converging beam is folded first by a flat mirror, and second, by a beam splitter which transmits to the guiding camera (not available during the 2005 run) and reflects to the coronagraph. The coronagraph's main component is the four quadrants phase mask, located in the focal plane of the telescope. The relay lens $L_1$ produces an image of the pupil in a plane where the Lyot stop eliminates the diffracted light. Finally, the imaging lens (L2 or L3) produces an image of the star or of the pupil on the science camera's sensor.

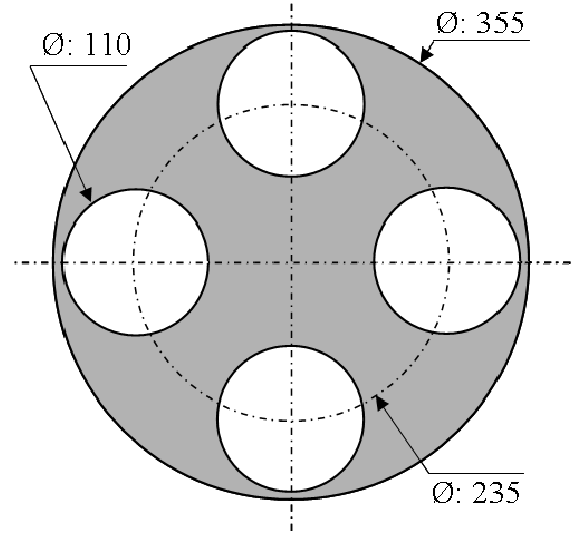

**Figure 4:** Geometry of the entrance pupil mask for the CORONA instrument. Dimensions are in millimeters.

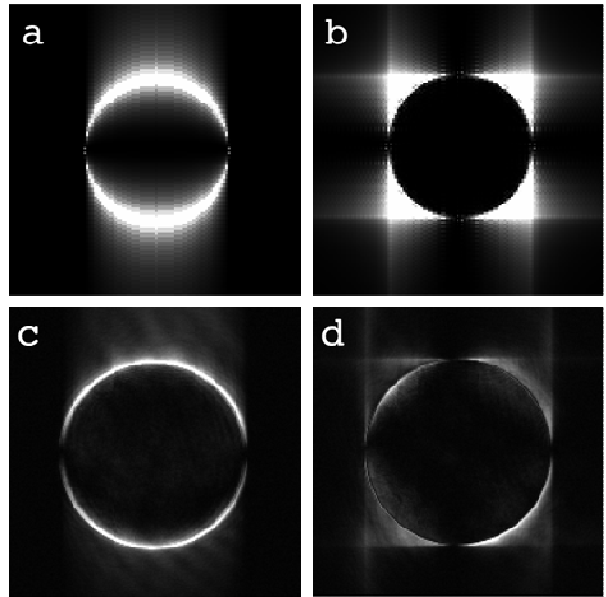

**Figure 5:** Coronagraphic images of the pupil plane (before the Lyot stop). Numerical simulations: **a:** the star image is on *one* phase knife; **b:** the star image is at the intersection of the *two* phase knives. Laboratory images: **c** and **d** correspond to the conditions of **a** and **b** respectively.

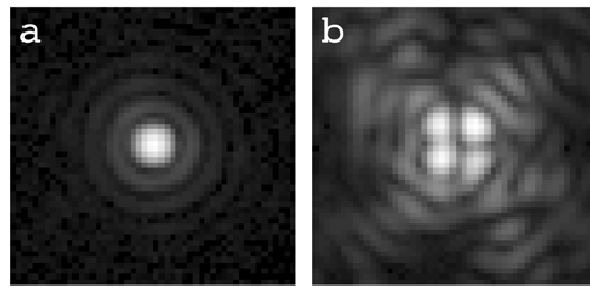

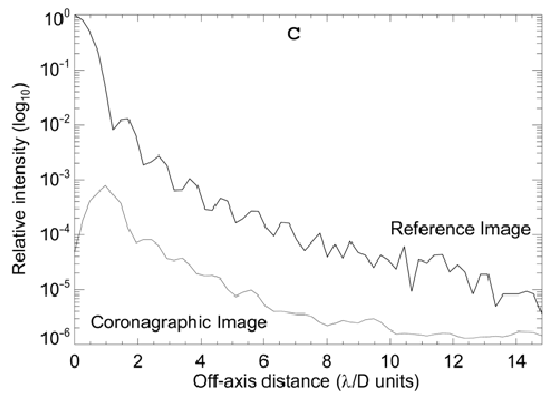

**Figure 6:** Laboratory experiment results obtained with a single circular entrance pupil. **a:** Reference image; **b**: Coronagraphic image (intensity x 1000); **c**: average radial profiles (logarithmic scale for the intensity axis).

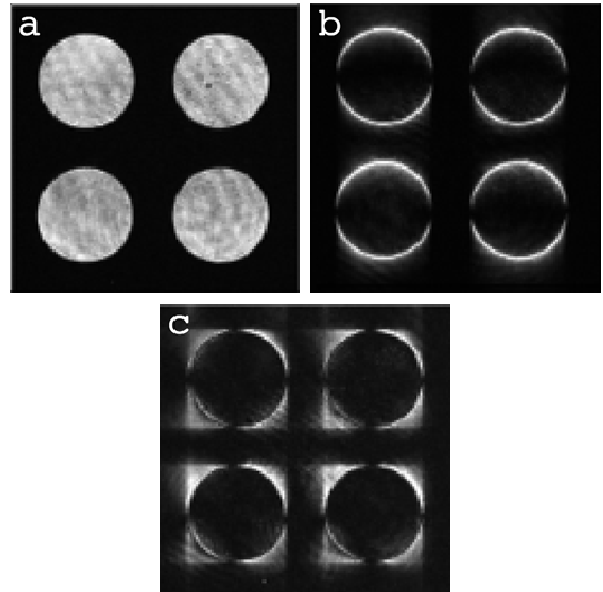

**Figure 7:** Laboratory pupil images. **a**: with the source off-axis; **b**: with the source image on *one* phase knife; **c**: with the source image at the intersection of both phase knives (coronagraphic pupil image).

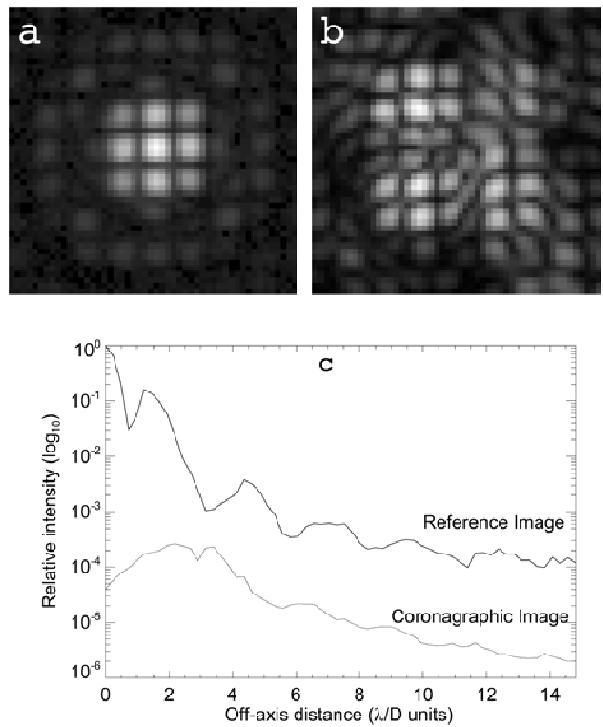

**Figure 8:** Laboratory experiment results obtained with the four sub-apertures pupil mask. **a**: reference image; **b**: coronagraphic image (intensity x 1000); **c**: average radial profiles (logarithmic scale for the intensity axis).

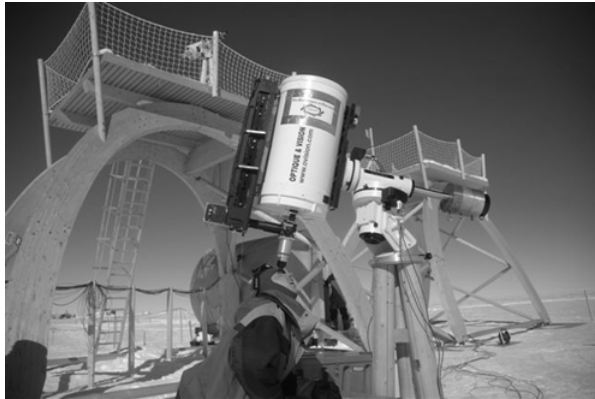

**Figure 9:** CORONA at Dome C in December 2005.

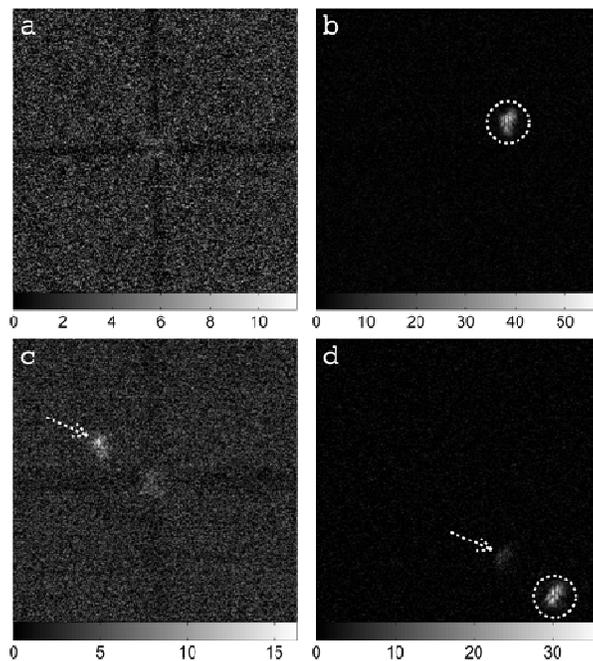

**Figure 10:** Best attenuated coronagraphic images: **a**: for HD 45348 (single star) and **c**: for HD 128620J (double star). For visual comparison, non-coronagraphic images of these two objects are also shown: **b**: for HD 45348 and **d**: for HD 128620J. On the non-coronagraphic images **b** and **d**, the dashed circles show the star which is on-axis on frames **a** and **c**. The dashed arrows on frames **c** and **d** show the companion.

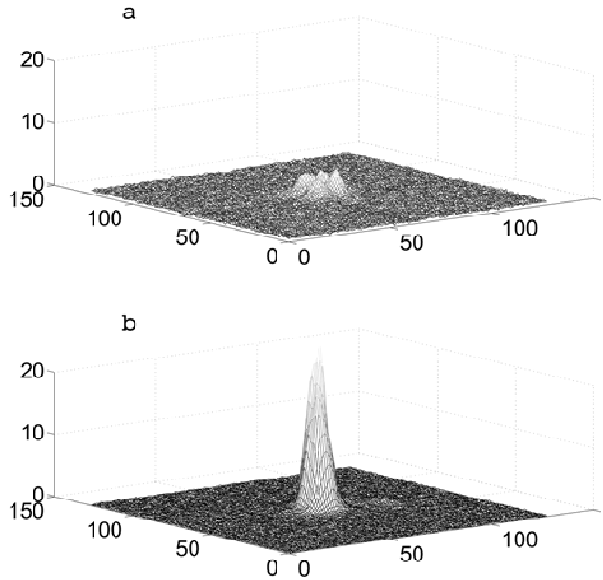

**Figure 11:** 3D intensity plots for long exposures of HD 45348. ***a***: coronagraphic image, ***b***: reference non-coronagraphic image.

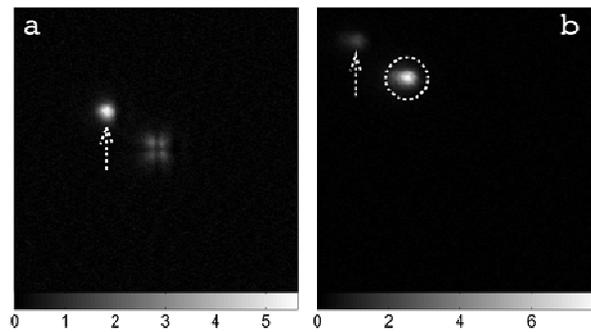

**Figure 12:** Long exposures of HD 128620J. ***a***: coronagraphic image, ***b***: non-coronagraphic image. On frame ***b***, the dashed circle shows the star which is on-axis on frame ***a***). The dashed arrows show the companion. The photometric scales are different on both frames.

## Tables

| Object | n | Max. rejection | Long exp. rejection | Max. extinction | Long exp. extinction |
|---|---|---|---|---|---|
| HD 45348 | 1122 | 15 | 11 | 17 | 24 |
| HD 128620J | 3740 | 9 | 10 | 9 | 14 |

**Table 1:** Summary of the results of CORONA's first daytime observation campaign at Dome C (December 2005). *n* is the total number of coronagraphic frames available. *Max. rejection* is the maximum rejection ratio on an individual snapshot. *Long exp. rejection* is the rejection ratio measured on a long exposure obtained from the 10 best coronagraphic images. *Max. extinction* is the maximum extinction ratio on an individual snapshot. *Long exp. extinction* is the extinction ratio measured on a long exposure obtained from the 10 best coronagraphic images.